\documentstyle[12pt,epsfig]{ioplppt} 
\def\las{\mathrel{\hbox{\rlap{\hbox{\lower4pt\hbox{$\sim$}}}\hbox{$<$}}}} 
\def\gas{\mathrel{\hbox{\rlap{\hbox{\lower4pt\hbox{$\sim$}}}\hbox{$>$}}}}

\def\etal{{\it et \thinspace al.}\ } 
\begin{document} 
\jl{2} 
\title{Electron Impact Excitation of Helium-like 
Oxygen up to n=4 levels including radiation damping} 
 
\author{Franck Delahaye and Anil K. Pradhan} 
 
\address{Department of Astronomy,  
Ohio State University \\ Columbus, Ohio, 
USA, 43210} 
 
\begin{abstract} 
 
The primary X-ray diagnostic lines in He-like ions 
are mainly excited by electron impact from the ground  
level to the $n = 2$ levels, but at high temperatures $n > 2$ levels  
are also excited. In order to describe the atomic processes more completely
collision strengths are computed for O~VII including for the first time
all of the following:
(i) relativistic fine structure, (ii) levels up to the $n=4$, and (iii)
radiation damping of autoionizing resonances.
The calculations are carried out using the  
Breit-Pauli R-matrix (BPRM) method with a 31-level eigenfunction expansion.  
Resonance structures in collision strengths are delineated in detail up to  
the $n = 4$ thresholds. For highly charged He-like ions radiation damping  
of autoionizing resonances is known to be significant. 
We investigate this effect in detail and find that while resonances  
are discernibly damped radiatively as the series limit $n \rightarrow \infty$ is
approached from below, the overall effect on effective cross sections and  
rate coefficients is found to be very small. Collision strengths for the
principal lines important in X-ray plasma diagnostics, w,x,y and z,
corresponding to the 4 transitions to
the ground level $1s^2 \ (^1S_0)
\longleftarrow 1s2p (^1P^o_1), 1s2p (^3P^o_2), 1s2p (^3P^o_1), 1s2s
(^3S_1)$, are explicitly shown. It is found that the effective
collision strength of the forbidden z-line is up to a factor of 4 higher
at T $< 10^6$ K than previous values. This is
likely to be of considerable importance in the diagnostics of
photoionized astrophysical
plasmas. Significant differences are also found with previous works for
several other transitions.
This work is carried out as part of the Iron Project-RmaX Network. 

\end{abstract}

\pacno{34.80.Kw} 
\maketitle 
 
\submitted 
 
\section{Introduction}

Helium-like ions provide the most important X-ray spectral diagnostics
in high temperature fusion and astrophysical plasmas.
The new generation of X-Ray satellites such as the Chandra X-Ray
Observatory and the X-Ray Multi-Mirror Mission-Newton provide
high resolution spectra of
different types of astronomical objects (e.g. Kaastra \etal 2000,
Porquet and Dubau 2000, Porquet \etal 2001). The high sensitivity of these 
observatories and the high quality of the spectra they produce requires
highly accurate atomic data for a precise interpretation. 
The aim of  the Iron Project-R-matrix calculations for X-ray  
spectroscopy (IP-RmaX) is to calculate extended sets of accurate collision  
strengths and rate coefficients for all ions of importance in X-Ray  
diagnostics. Among previous works, the electron impact excitation of
Helium like oxygen was
previously considered by Pradhan \etal(1981a,b) in the distorted wave  
and close coupling approximations for transitions up to the $n=2$ levels.
 Sampson \etal (1983)
and Zhang and Sampson (1987) used the Coulomb-Born approximation with
exchange, intermediate coupling, and some resonances effects to  
obtain collision strengths for Helium-like ions, with atomic number
Z spanning a large  
range of values ($4<Z<74$). Kingston and Tayal (1983a,b) calculated the  
collision strength for two 
transitions, from the ground state to $2 ^3S_1$ and to $2 ^3P^o_1$, using  
the close coupling R-matrix (RM) method, and derived the corresponding
effective collision  
strengths. Both, the Pradhan \etal and Kingston and Tayal calculations 
were in LS coupling. The present work aims at generating a more complete 
dataset of  
high reliability for O VII, including all important effects for highly  
charged ions such as relativistic effects, radiation damping, and  
resonances in higher complexes up to $n=4$.

The method and computations are summerized in section 2.
Results for the collision strengths and important issues  
are discussed in section 3, and the present results for the effective
(Maxwellian averaged) collision strengths are
compared with previous calculations. The main conclusions are given in
section 4, together with an estimate of accuracy of the results.

\section{Method and Computations\protect\\} 
 
The collisional calculation in the present work has been carried out using 
the Breit-Pauli R-matrix (BPRM) method as used in the Iron Project (IP) and
utilised in a number of previous publications.
The aims and methods of the IP are presented in Hummer \etal(1993).
We briefly summarise the main features of the method and calculations.

In the coupled channel or close coupling (CC) approximation
the wave function expansion,
$\Psi(E)$, for a total spin and angular symmetry  $SL\pi$ or $J\pi$,
of the (N+1) electron system
is represented in terms of the target ion states as:

\begin{equation}
\Psi(E) = A \sum_{i} \chi_{i}\theta_{i} + \sum_{j} c_{j} \Phi_{j},
\end{equation}

\noindent
where $\chi_{i}$ is the target ion wave function in a specific state
$S_iL_i\pi_i$ or level $J_i\pi_i$, and $\theta_{i}$ is the wave function
for
the (N+1)$^{th}$ electron in a channel labeled as
$S_iL_i(J_i)\pi_i \ k_{i}^{2}\ell_i(SL\pi) \ [J\pi]$; $k_{i}^{2}$ is the
incident kinetic energy. In the second sum the $\Phi_j$'s are
correlation
wave functions of the (N+1) electron system that (a) compensate for the
orthogonality conditions between the continuum and the bound orbitals,
and (b) represent additional short-range correlations that are often of
crucial importance in scattering and radiative CC calculations for each
symmetry. The $\Phi_j$'s are also referred to as ``bound channels", as
opposed to the continuum or ``free" channels in the first sum over the
target states. In the relativistic BPRM calculations the set of
${SL\pi}$
are recoupled in an intermediate (pair) coupling scheme 
to obtain (e + ion) states  with total $J\pi$, followed by
diagonalisation of the (N+1)-electron Hamiltonian. Details of the
diagonalization and the R-matrix method are given in many previous works
(e.g. Berrington \etal 1995).

The target expansion for the close coupling calculations consists of 31  
fine-structure levels arising from the 19 LS terms with principal quantum  
number $n\leq 4$. The target eigenfunctions were developed using the
SUPERSTRUCTURE program
(Eissner \etal1974) with a version due to Nussbaumer and Storey (1978).
The full expansion, together with the scaling  
factors in the {\it Thomas-Fermi} potential employed in SUPERSTRUCTURE,
are given at the end of table 1.
 
In order to estimate the quality of the target wavefunction expansion,  
we compare the energy levels with those from the {\it National Institue
for Standards and Technology} (NIST) in table 1. A better
criterion for the accuracy of the wavefunctions is the accuracy
of the oscillator strengths for transitions in the target ion. In table 2 we 
compare the {\it gf}-values with the evaluated compilation from NIST for a 
number of dipole transitions in O~VII. For the energies the agreement with
the NIST values is found to be very good, within 0.05\% for all levels.
The oscillator strengths agree well within
10\% (however, for some of the values given by NIST, the estimated accuracy  
is 30\%). Another accuracy criterion is the level of agreement between the 
oscillator strengths in the length and the velocity formulations,
which we also find to be a few percent for all transitions. The Einstein
A-values are also presented in table 2 to enable ready application of
the present collisional data in radiative-collisional models for
spectral diagnostics (e.g. Porquet and Dubau 2000, Porquet \etal 2001).
 
\begin{table}
\centering
\noindent{Table 1: Energy levels compared  
          with observed values (in Rydbergs).}
\newline 
\begin{tabular}{llrr} 
\hline 
\noalign{\smallskip} 
 Levels & & $E_{calc}$ 
&   $E_{obs}$  \\ 
            \noalign{\smallskip} 
            \hline 
            \noalign{\smallskip} 
$1s^2 $ & $^1S{_0}$ & 0.0000 & 0.0000 \\ 
$1s2s$ & $^3S{_1}$ & 41.2438 & 41.2315 \\ 
$1s2p$ & $^3P{_0}^o$ & 41.7933 & 41.7872 \\ 
$1s2p$ & $^3P{_1}^o$ & 41.7942 & 41.7877 \\ 
$1s2p$ & $^3P{_2}^o$ & 41.7997 & 41.7928 \\ 
$1s2s$ & $^1S{_0}$ & 41.8074 & 41.8124 \\ 
$1s2p$ & $^1P{_1}^o$ & 42.2100 & 42.1844 \\ 
$1s3s$ & $^3S{_1}$ & 48.6577 & 48.6509 \\ 
$1s3p$ & $^3P{_0}^o$ & 48.8114 & 48.8044 \\ 
$1s3p$ & $^3P{_1}^o$ & 48.8116 & 48.8044 \\ 
$1s3p$ & $^3P{_2}^o$ & 48.8132 & 48.8044 \\ 
$1s3s$ & $^1S{_0}$ & 48.8217 & 48.8112 \\ 
$1s3d$ & $^3D{_1}$ & 48.8930 & 48.8837 \\ 
$1s3d$ & $^3D{_2}$ & 48.8931 & 48.8842 \\ 
$1s3d$ & $^3D{_3}$ & 48.8935 & 48.8843 \\ 
$1s3d$ & $^1D{_2}$ & 48.8971 & 48.8937 \\ 
$1s3p$ & $^1P{_1}^o$ & 48.9281 & 48.9218 \\ 
$1s4s$ & $^3S{_1}$ & 51.1813 & 51.1798 \\ 
$1s4p$ & $^3P{_0}^o$ & 51.2436 & 51.2360 \\ 
$1s4p$ & $^3P{_1}^o$ & 51.2437 & 51.2360 \\ 
$1s4p$ & $^3P{_2}^o$ & 51.2444 & 51.2360 \\ 
$1s4s$ & $^1S{_0}$ & 51.2475 & 51.2410 \\ 
$1s4d$ & $^3D{_1}$ & 51.2767 & 51.2675 \\ 
$1s4d$ & $^3D{_2}$ & 51.2767 & 51.2600 \\ 
$1s4d$ & $^3D{_3}$ & 51.2769 & 51.2720 \\ 
$1s4f$ & $^3F{_2}^o$ & 51.2786 & 51.2698 \\ 
$1s4f$ & $^3F{_3}^o$ & 51.2785 & 51.2698 \\ 
$1s4f$ & $^3F{_4}^o$ & 51.2787 & 51.2698 \\ 
$1s4d$ & $^1D{_2}$ & 51.2790 & 51.2739 \\ 
$1s4f$ & $^1F{_3}^o$ & 51.2787 & 51.2755 \\ 
$1s4p$ & $^1P{_1}^o$ & 51.2916 & 51.2870 \\ 
\noalign{\smallskip} 
\hline\\ 
\end{tabular} 
\\
Spectroscopic configurations: $1s^2$, 1s2s, 1s2p 
1s3s, 1s3p, 1s3d, 1s4s, 1s4p, 1s4d, 1s4f.\\
Correlation configurations: $2s^2$, $2p^2$, 2s2p 
2s3s, 2s3p, 2s3d, 2s4s, 2s4p, 2s4d, 2s4f.\\
Scalling factors: $\lambda_{1s}$ = 0.9932, 
$\lambda_{2s}$ = 1.0759, $\lambda_{2p}$ = 0.9217, $\lambda_{3s}$ = 1.0306,\\ 
$\lambda_{3p}$ = 0.9023, $\lambda_{3d}$ = 0.9547, $\lambda_{4s}$ = 1.0182,
$\lambda_{4p}$ = 0.9038, $\lambda_{4d}$ = 0.9512,$\lambda_{4f}$ = 1.0200.\\
\end{table}

 
\begin{table} 
\noindent{Table 2: Comparison of Aij and g*f 
           values compiled by NIST and thoses 
           obtained with SUPERSTRUCTURE (calc).\\ \\} 
\begin{tabular}{ccrrrrrr} 
\hline 
\noalign{\smallskip} 
 $Conf_i-Conf_k$ &$Term_i-Term_k$ & $E_i\ (Ryd)$ 
&   $E_k\ (Ryd)$ & $A_{ki}^{NIST}\ (s)$ & $A_{ki}^{cal}\ (s)$ 
& $g*f_{NIST}$ & $g*f_{cal}$ \\ 
            \noalign{\smallskip} 
            \hline 
            \noalign{\smallskip} 
$ 1s2-1s2p$& $ ^1S_0$ - $^1P^o_1$&   0.0000 &  42.1844 & 3.309e+12 &3.403e+12 &6
.945e-01 &7.142e-01 \\ 
$ 1s2-1s3p$&$ ^1S_0$ - $^1P^o_1 $&   0.0000 &  48.9218 & 9.365e+11 &1.004e+12 &1
.462e-01 &1.566e-01 \\ 
$ 1s2s-1s2p$&$ ^3S_1$ - $^3P^o_0$&  41.2315 &  41.7872 & 7.797e+07 &8.058e+07 &3
.143e-02 &3.249e-02 \\ 
$ 1s2s-1s2p$&$ ^3S_1$ - $^3P^o_1$&  41.2315 &  41.7877 & 7.820e+07 &8.083e+07 &9
.441e-02 &9.757e-02 \\ 
$ 1s2s-1s2p$&$ ^3S_1$ - $^3P^o_2$&  41.2315 &  41.7928 & 8.033e+07 &8.309e+07 &1
.587e-01 &1.642e-01 \\ 
$ 1s2p-1s3s$&$ ^3P^{o}_0$ - $ ^3S_1$&  41.7872 &  48.6509 & 2.505e+09 &2.237e+09
 &1.986e-02 &1.774e-02 
\\ 
$ 1s2p-1s3d$&$ ^3P^{o}_0$ - $ ^3D_1$&  41.7872 &  48.8837 & 8.982e+10 &8.927e+10
 &6.662e-01 &6.621e-01 
\\ 
$ 1s2p-1s3s$&$ ^3P^{o}_1$ - $ ^3S_1$&  41.7877 &  48.6509 & 7.512e+09 &6.739e+09
 &5.957e-02 &5.344e-02 
\\ 
$ 1s2p-1s3d$&$ ^3P^{o}_1$ - $ ^3D_1$&  41.7877 &  48.8837 & 6.735e+10 &6.699e+10
 &4.996e-01 &4.969e-01 
\\ 
$ 1s2p-1s3d$&$ ^3P^{o}_1$ - $ ^3D_2$&  41.7877 &  48.8842 & 1.213e+11 &1.205e+11
 &1.499e+00 &1.490e+00 
\\ 
$ 1s2p-1s3s$&$ ^3P^{o}_2$ - $ ^3S_1$&  41.7928 &  48.6509 & 1.249e+10 &1.131e+10
 &9.917e-02 &8.983e-02 
\\ 
$ 1s2p-1s3d$&$ ^3P^{o}_2$ - $ ^3D_1$&  41.7928 &  48.8837 & 4.481e+09 &4.466e+09
 &3.328e-02 &3.318e-02 
\\ 
$ 1s2p-1s3d$&$ ^3P^{o}_2$ - $ ^3D_2$&  41.7928 &  48.8842 & 4.033e+10 &4.012e+10
 &4.992e-01 &4.966e-01 
\\ 
$ 1s2p-1s3d$&$ ^3P^{o}_2$ - $ ^3D_3$&  41.7928 &  48.8843 & 1.613e+11 &1.608e+11
 &2.795e+00 &2.787e+00 
\\ 
$ 1s2s-1s2p$&$ ^1S_0$ - $^1P^o_1$&  41.8124 &  42.1844 & 2.514e+07 &2.509e+07 &6
.786e-02 &6.773e-02 \\ 
$ 1s2s-1s3p$&$ ^1S_0$ - $^1P^o_1$&  41.8124 &  48.9218 & 5.055e+10 &5.209e+10 &3
.735e-01 &3.849e-01 \\ 
$ 1s2s-1s3p$&$ ^1S_0$ - $^1P^o_1$&  41.8124 &  48.9218 & 5.055e+10 &5.209e+10 &3
.735e-01 &3.849e-01 \\ 
$ 1s2p-1s3s$&$ ^1P^{o}_1$ - $ ^1S_0$&  42.1844 &  48.8112 & 2.008e+10 &2.223e+10
 &5.692e-02 &6.303e-02 
\\ 
$ 1s2p-1s3d$&$ ^1P^{o}_1$ - $ ^1D_2$&  42.1844 &  48.8937 & 1.523e+11 &1.540e+11
 &2.106e+00 &2.130e+00 
\\ 
$ 1s3p-1s3d$&$ ^3P^{o}_0$ - $ ^3D_1$&  48.8044 &  48.8837 & 6.114e+05 &6.200e+05
 &3.632e-02 &3.684e-02 
\\ 
$ 1s3p-1s3d$&$ ^3P^{o}_1$ - $ ^3D_1$&  48.8044 &  48.8837 & 4.585e+05 &4.649e+05
 &2.723e-02 &2.762e-02 
\\ 
$ 1s3p-1s3d$&$ ^3P^{o}_1$ - $ ^3D_2$&  48.8044 &  48.8842 & 8.426e+05 &8.535e+05
 &8.236e-02 &8.338e-02 
\\ 
$ 1s3p-1s3d$&$ ^3P^{o}_2$ - $ ^3D_1$&  48.8044 &  48.8837 & 3.057e+04 &3.099e+04
 &1.816e-03 &1.841e-03 
\\ 
$ 1s3p-1s3d$&$ ^3P^{o}_2$ - $ ^3D_2$&  48.8044 &  48.8842 & 2.809e+05 &2.841e+05
 &2.746e-02 &2.775e-02 
\\ 
$ 1s3p-1s3d$&$ ^3P^{o}_2$ - $ ^3D_3$&  48.8044 &  48.8843 & 1.127e+06 &1.143e+06
 &1.539e-01 &1.560e-01 
\\ 
$ 1s3s-1s3p$&$ ^1S_0$ - $^1P^o_1$&  48.8112 &  48.9218 & 3.864e+06 &3.958e+06 &1
.180e-01 &1.210e-01 \\ 
$ 1s3d-1s3p$&$ ^1D_2$ - $^1P^o_1$&  48.8937 &  48.9218 & 7.410e+04 &8.082e+04 &3
.508e-02 &3.832e-02 \\ 
\noalign{\smallskip} 
\hline 
\end{tabular} 
\end{table} 
 
 
 The 19 LS terms are recoupled in the relativistic BPRM calculations 
into the corresponding 31 fine structure 
levels up to the  $n=4$ complex using the routine
RECUPD that performs intermediate coupling
operations including the one-body Breit-Pauli operators (Hummer et. al.
1993). The reconstructed target eigenfunctions and the
resulting target energies reproduce to $10^{-5}$ Ryd the results from  
SUPERSTRUCTURE, verifying that the algebraic operations have been
carried out self-consistently and without loss of accuracy. 
The collision strengths have been calculated for electron  
energies $0 \leq E \leq 200$ Ryd. This wide energy range ensures a good
coverage of the region where resonances up to the $n=4$ complex are important,  
as well as the higher energy region where no resonance have been included
(all channels are open) but where  
the background collision strengths still make a significant contribution
to the Maxwellian averaged rate coefficient for electron temperatures
of interest.
 
 The inner region R-matrix basis set included 50 orbitals per angular
momentum.
Because of the importance of the near threshold resonances in the Maxwellian  
average rate coefficient, careful attention has been devoted to the  
resolution and a precise mesh has been chosen. A mesh of $10^{-4}$ Ryd was  
selected for the region where resonances are important, and a coarser mesh for   
the region where all channel are open. 
We included the contribution to the collision strengths from all symmetries
with total angular momentum J and both odd and even parities,  
$J\pi \leq (\frac{35}{2})^{o,e}$. The contribution of higher partial waves
 was included  
using the Coulomb-Bethe approximation via the `top-up' facility 
in the asymptotic region program STGF of the R-matrix package
(Burke and Seaton 1986; modified by W. Eissner and G.X. Chen).

\section{Results and discussion\protect\\} 

In figures 1 and 2 we present the collision strengths for transitions from  
the ground state to levels in the $n=2$ complex, and to levels in the
complex $n=3$ respectively. The high resolution of the calculations with
a large number of points allow us to resolve clearly all the resonances up to
the last threshold in the n=4 complex. We delineate the Rydberg series
converging to the different series limits in all three complexes.
In both figure 1 and figure 2, the identification of the Rydberg series
converging to n=3 and $n=4$ thresholds has been marked.
The doubly excited (e +  O~VII) $\rightarrow$ O~VI resonance complexes, 
KMM, KMN etc.. converging  
towards the different  $n=3$ and $n=4$ levels are clearly resolved. 
We can anticipate from figure 2 that, for some transitions, the low magnitude of  
the background and the high density of the resonances will make the  
contribution of these resonances converging to the $n=4$ complex very  
important and dominant. This is confirmed by our work (discussed
later).
 
\begin{figure} 
\psfig{figure=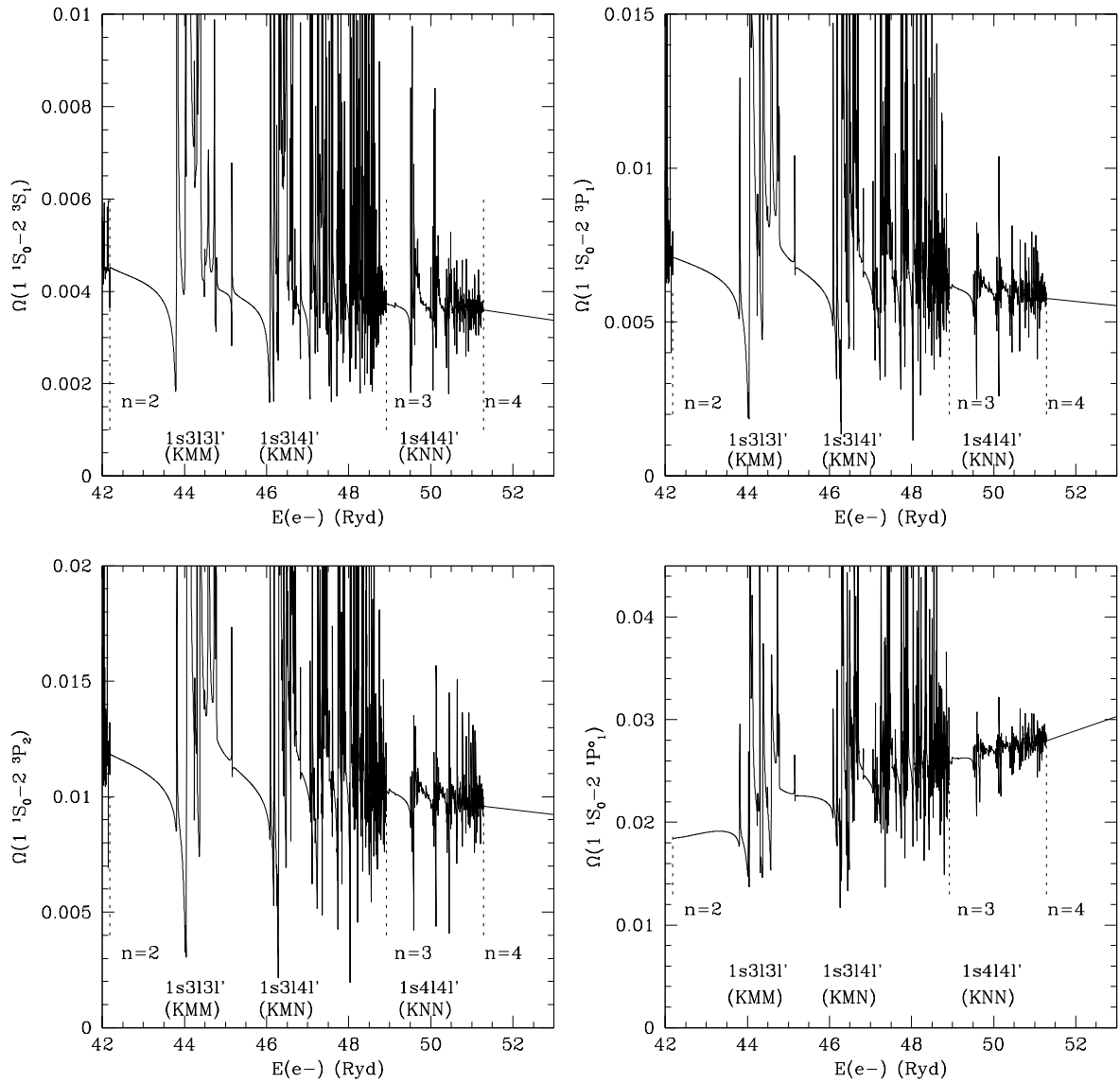,height=17.0cm,width=17.0cm} 
\vspace*{-1cm} 
\caption{Collision strengths for transition for the principal lines, z, 
\ x, y and w (from ground state $1s^2$ $^1S_0$ to $2 ^3S_1,\ 2 ^3P^o_1, 
\ 2 ^3P^o_2,\ 2 ^1P^o_1$).} 
\end{figure} 
 
\begin{figure} 
\psfig{figure=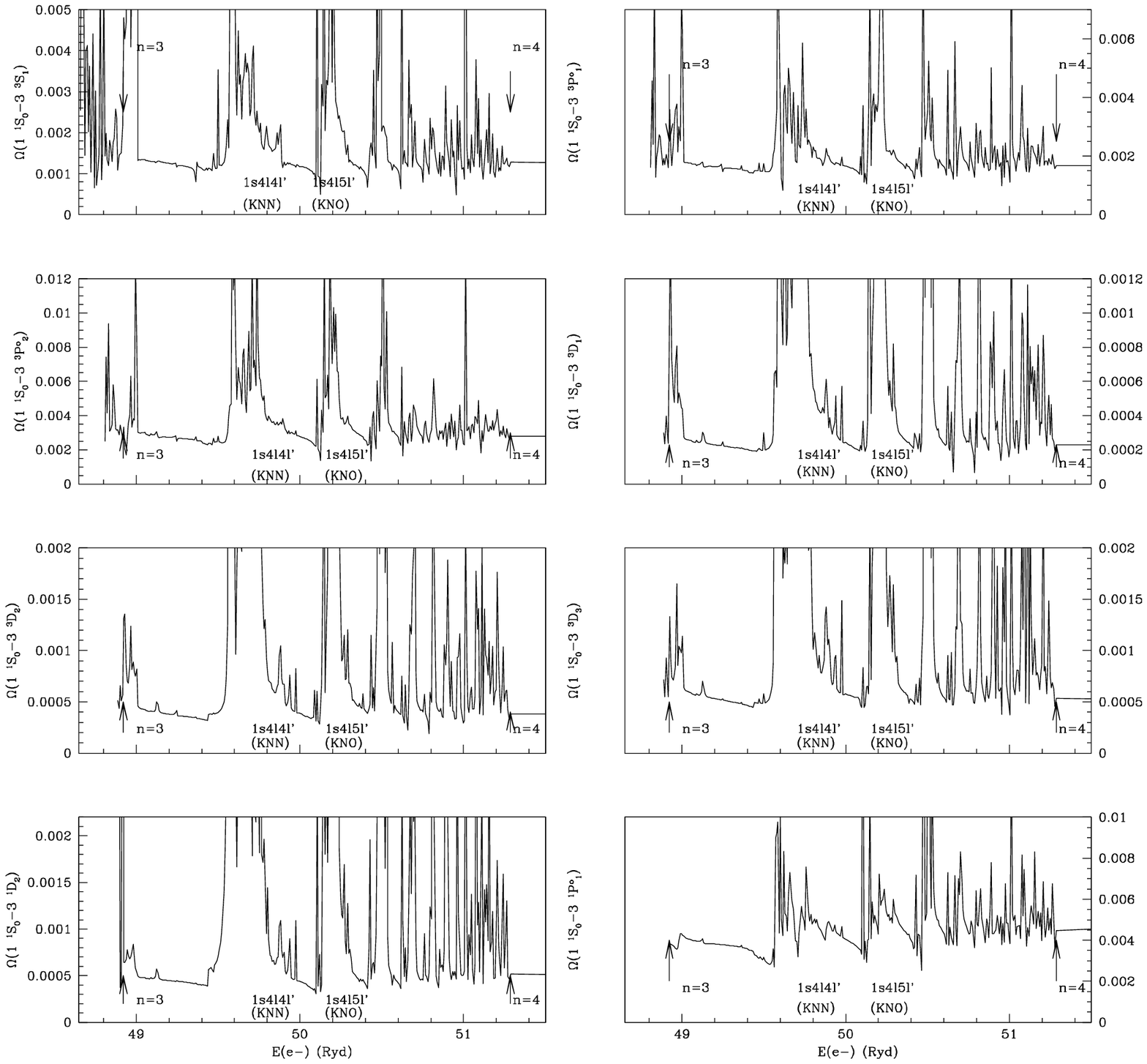,height=17.0cm,width=17.0cm} 
\vspace*{-1cm} 
\caption{ Collision strength for transition 
from the ground state to the complex with principal quantum 
number n=3. 
$1s^2\ ^1S_0 \rightarrow  3 ^3S_1,\ 3 ^3P^o_1,\ 3 ^3P^o_2,$ 
$\ 3 ^3D_1,\ 3 ^3D_2,\ 3 ^3D_3,\ 3 ^1D_2,\ 3 ^1P^o_1$} 
\end{figure}

\subsection{Radiation Damping} 
It has been previously shown (Presnyakov and Urnov 1979, Pradhan 1981,
Pradhan and Seaton 1985), that radiation damping may have significant 
effect on the resonances in collision strengths for highly charged ions
since the radiative decay rates
may be large and may compete with autoionization rates, i.e. the effect
of dielectronic recombination on electron impact excitation.
We studied in detail the radiation damping effect of dielectronic
recombination, on resonance structures, collision strengths, and
rate coefficients.
In figures 3 and 4, we present  the Rydberg series converging to the $n=2$
and $n=3$ levels coupled to the ground state $1^1S_0$ via strong dipole
transitions, $2^1P^o_1$  
(figure3) and $3^1P^o_1$ (figure4). Since the autoionization rates decrease  
as $n^{-3}$, and the radiative rate remains constant, radiation damping  
increases with $n$ and the  resonances are wiped out as the series
limit is reached. We illustrate the effect for one total (e + ion) symmetry  
$J\pi = 1^e$. It can be seen how effective the diminishing of resonances  
is as we approach the threshold. The overall effect of damping, on the averaged
collision strength, can be up to a factor two (figure5). However, the region  
where the effect is important is very small, just below the threshold of  
convergence. It is called the 'quantum defect region', since we use the Bell  
and Seaton (1985) (see also Pradhan and Seaton 1985) multi-channel quantum defect theory in this region.  
In figures 4 and 5 this region corresponds to $\Delta E=0.36$ Ryd  
($\nu_{min} = 10$). Overall however we find that 
for O~VII the effective collision strengths and the rate 
coefficients are not significantly affected (figure 6 and 7). Indeed, in 
figure 6, damped and undamped effective collision strength $\Upsilon$ 
curves are indistinguishable (solid line). 

Figures 4 and 5 show that below threshold the collision strength is constant. 
This is due to the high value of the effective quantum number reached in
our calculations ($\nu \approx 100$), sufficient to illustrate
graphically the effect of radiation damping up to the region
where the resonances are almost completely damped.
In order to resolve all resonances converging to the different thresholds of
interest, we used an $\nu$-mesh with about 1000 points for each
interval $(\nu\ ,\ \nu +1)$. In the quantum defect region the Coulomb
potential dominates the scattering process.
In order to directly demonstrate the effect of radiation damping in figures
4 and 5, we
show that the collision strengths converge toward the background value
calculated neglecting the long-range non-dipole potentials.
However, just above the threshold the potential is not only Coulombic
but multipole contributions are also important. In order to isolate
the effect we switched-off the multipole contributions so
that there is continuity across the threshold and only the radiation  
damping effect is illustrated.

\begin{figure} 
\centering 
\psfig{figure=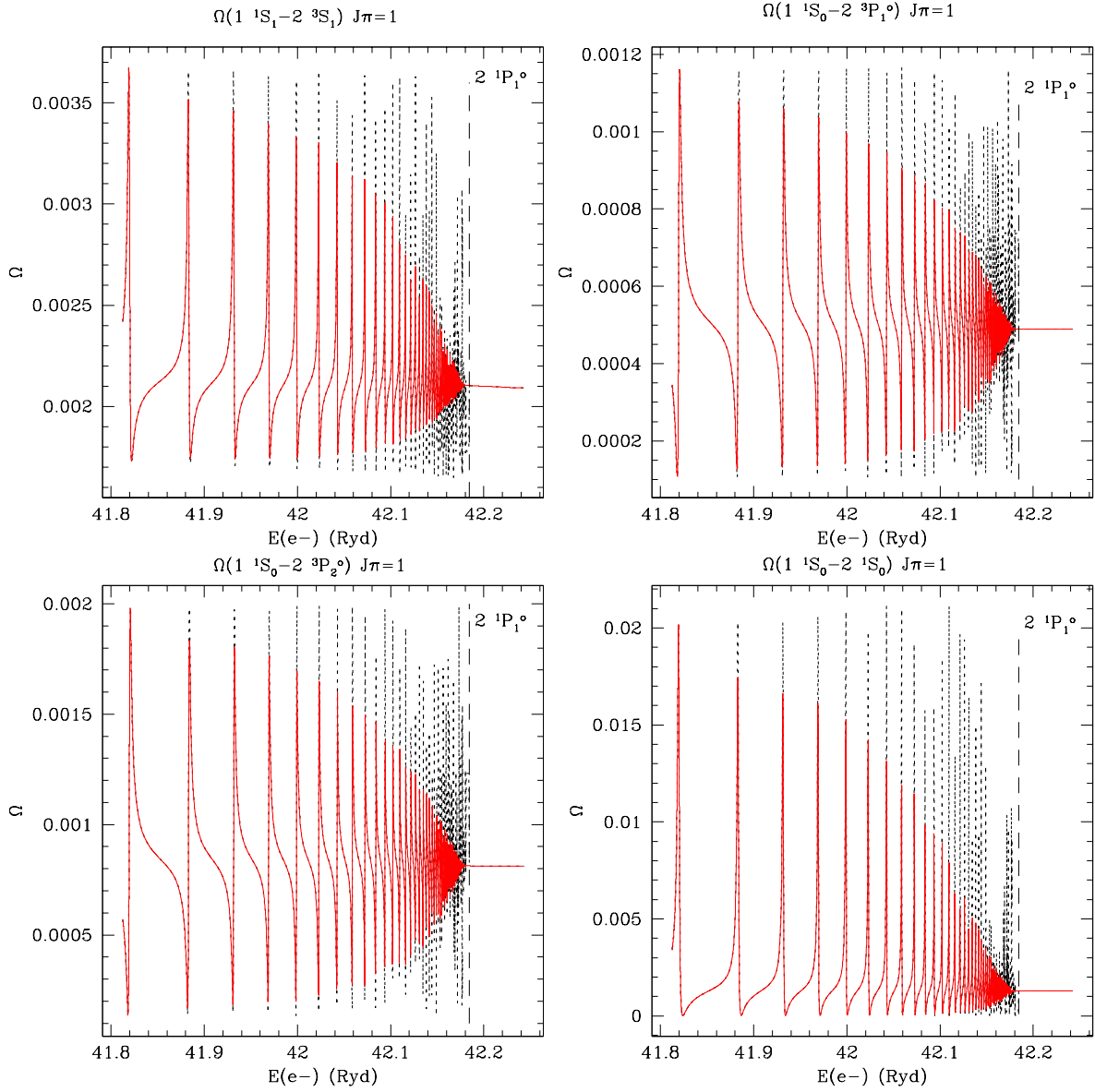,height=17.0cm,width=17.0cm} 
\vspace*{-1cm} 
\caption{Partial Collision strength for transition from 
ground state $1s^2$ $^1S_0$ to $2 ^3S_1,\ 2 ^3P^o_1,\ 2 ^3P^o_2,\ 2 ^1S_0$. 
dashe line: resonance for $J\pi=1_{even}$ converging to $2 ^1P^o_1$; 
solid line: resonance for $J\pi=1_even$ converging to $2 ^1P^o_1$ damped 
by recombination.} 
\end{figure} 
 
\begin{figure} 
\centering 
\psfig{figure=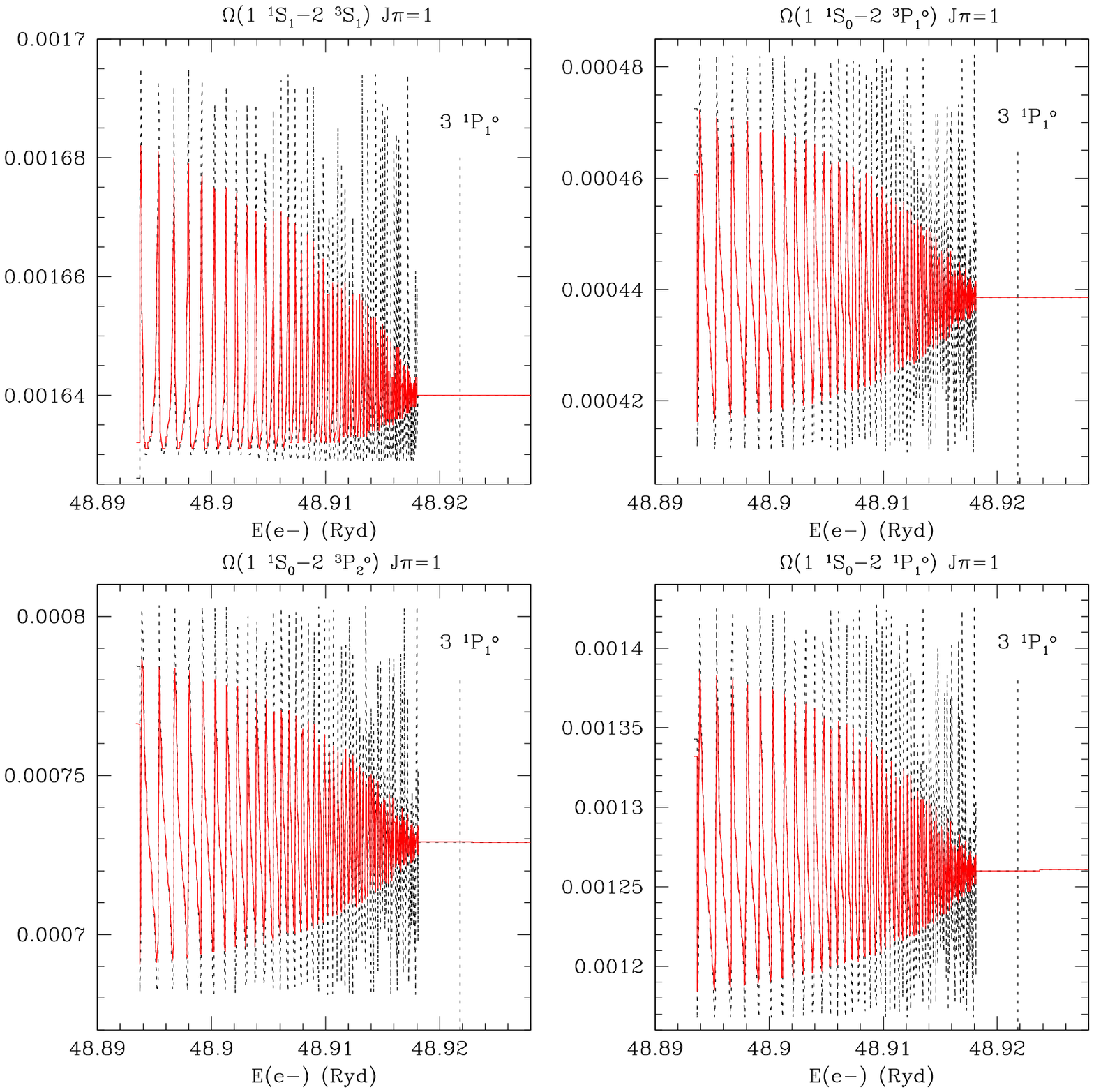,height=17.0cm,width=17.0cm} 
\vspace*{-1cm} 
\caption{Partial Collision strength for transition from 
ground state $1s^2$$^1S_0$ to $3 ^3S_1,\ 3 ^3P^o_1,\ 3 ^3P^o_2,\ 3 ^1P^o_1$ 
dashe line: resonance for $J\pi=1_{even}$ converging to $3 ^1P^o_1$; 
solid line: resonance for $J\pi=1_{even}$ converging to $3 ^1P^o_1$ damped 
by recombination.} 
\end{figure} 
 
\begin{figure} 
\centering 
\psfig{figure=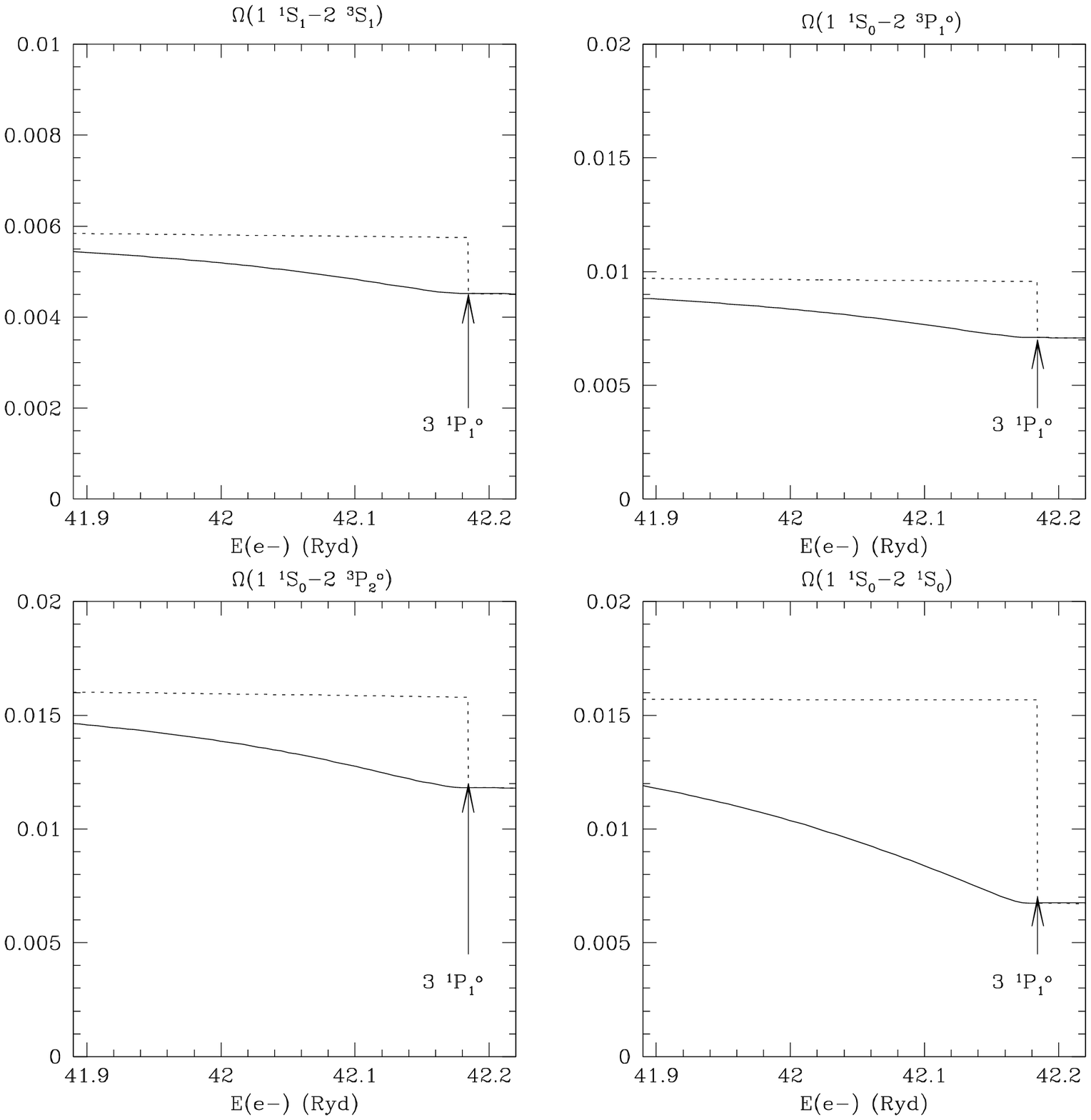,height=17.0cm,width=17.0cm} 
\vspace*{-1cm} 
\caption{Total averaged Collision strength for transition from 
ground state $1s^2$$^1S_0$ to $2 ^3S_1,\ 2 ^3P^o_1,\ 2 ^3P^o_2,\ 2 ^1S_0$ 
between  $2 ^1S_0$ and $2 ^1P^o_1$ thresholds. 
dashed line: No damping; solid line: Damping effect} 
\end{figure}

\subsection{Effective Collision Strengths}

 The Maxwellian averaged collision strengths

\begin{equation}
\Upsilon (T) = \int ^{\infty}_{0}
\Omega_{ij}(\epsilon_{j}) e^{-\epsilon_{j}/kT} d(\epsilon_{j}/kT),
\end{equation}

 have been computed for all transitions among levels up to the $n=4$.
As has been shown for He-like Fe XXV  (Kimura et
al 1999, 2000, and Machado-Pelaez et. al. 2001),
the resonances arising from the complex $n=N+1$
have a strong effect on transitions to the complex $n=N$.
Along with the full calculation including all the 31 fine structure states up
to $n=4$, we considered a smaller target model, including all 17 levels
up to $n=3$ to compare the effect of the resonances.
 figures 6, 7 and 8 show the effective collision strengths for 
transitions from the ground state to the $n=2$ levels, as well as
transitions among levels of the $n=2$ complex, and transitions from the 
ground to levels of the $n=3$ complex. We find that for some
transitions the
contribution from the higher $n=4$ resonances gives rise to a factor of two  
increase in the
effective collision strengths at the temperature of maximum abundance of O~VII
($T \approx 2 \times \ 10^{6}$).

\begin{figure} 
\centering 
\psfig{figure=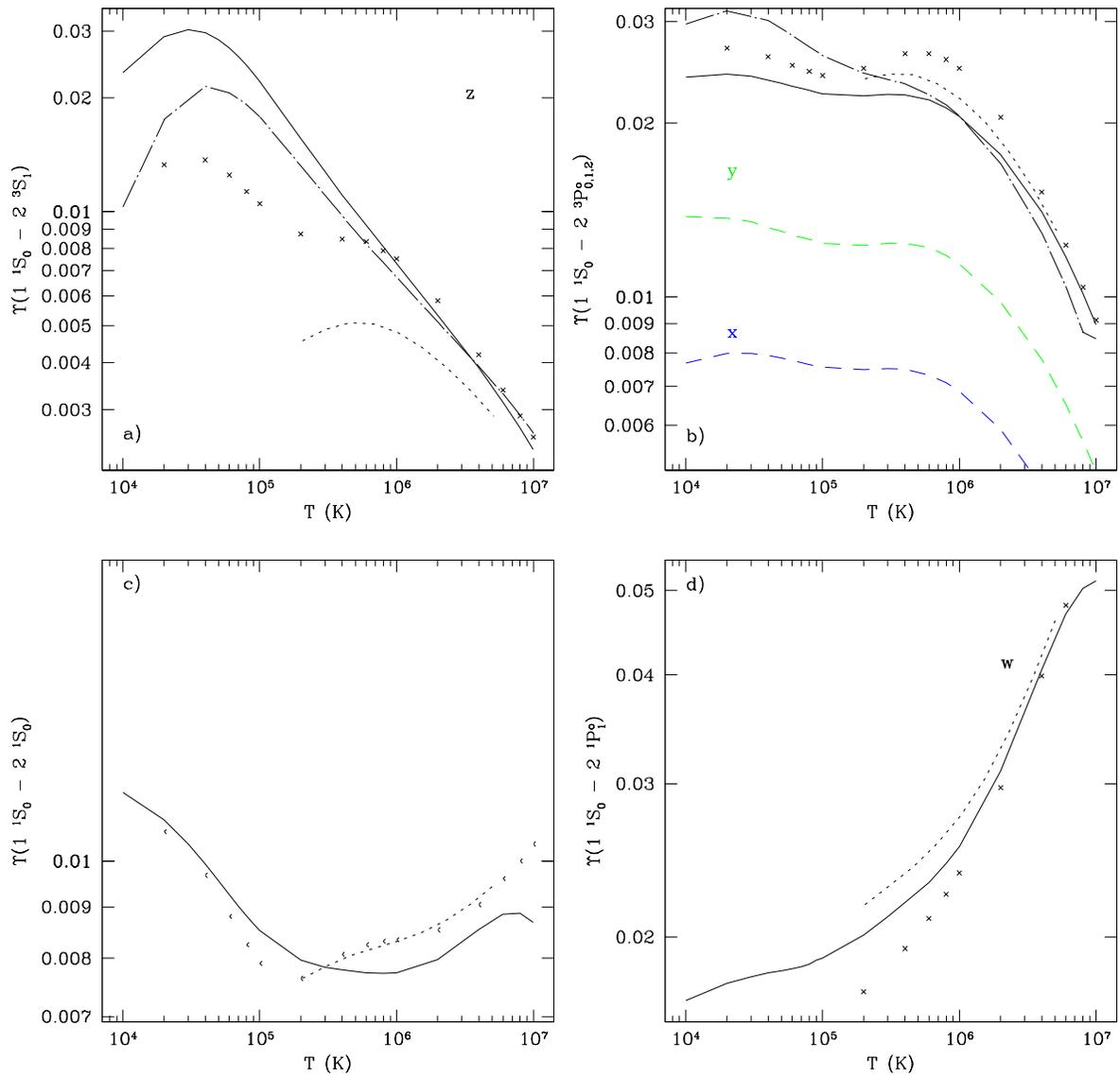,height=17.0cm,width=17.0cm} 
\caption{Effective Collision Strengths for the principal lines (z, x, y, 
w). 
solid line: Present work (long dashes on 6b: x and y; solid =(x+y)) 
dot-dashe line: Kingston and Tayal (1983)
short-dashe line: Zhang and Sampson (1987) 
Crosses: Pradhan \etal (1981) 
Note: The two curves with and without radiation damping effect are 
indistinguishable (solid line).} 
\end{figure} 
 
\begin{figure} 
\centering 
\psfig{figure=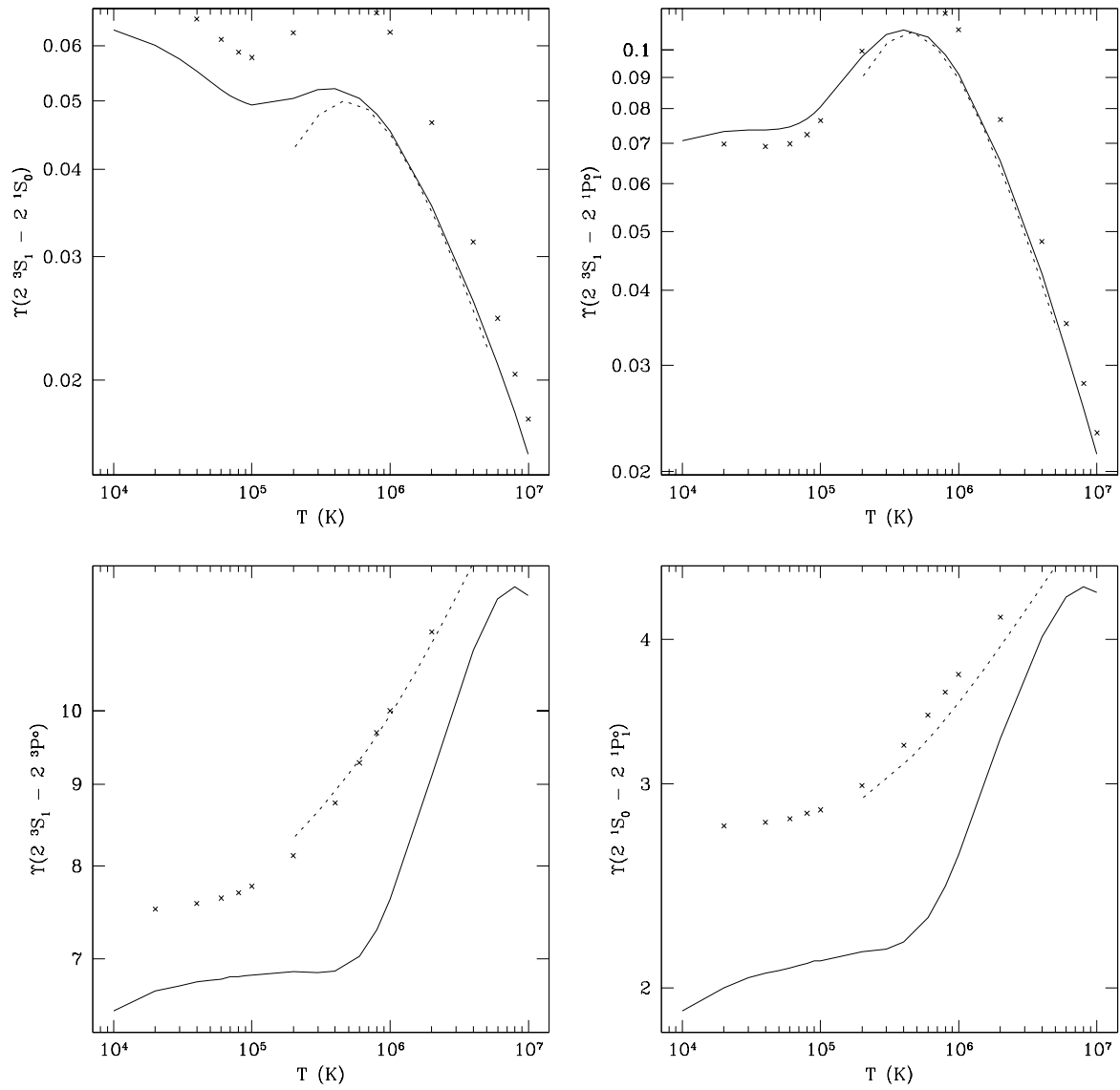,height=17.0cm,width=17.0cm} 
\caption{Effective Collision Strengths for transitions within the $n=2$
complex. 
solid line: Present work 
dashed line: Zhang and Sampson (1987) 
crosses: Pradhan \etal (1981)} 
\end{figure} 
\begin{figure} 
\centering 
\psfig{figure=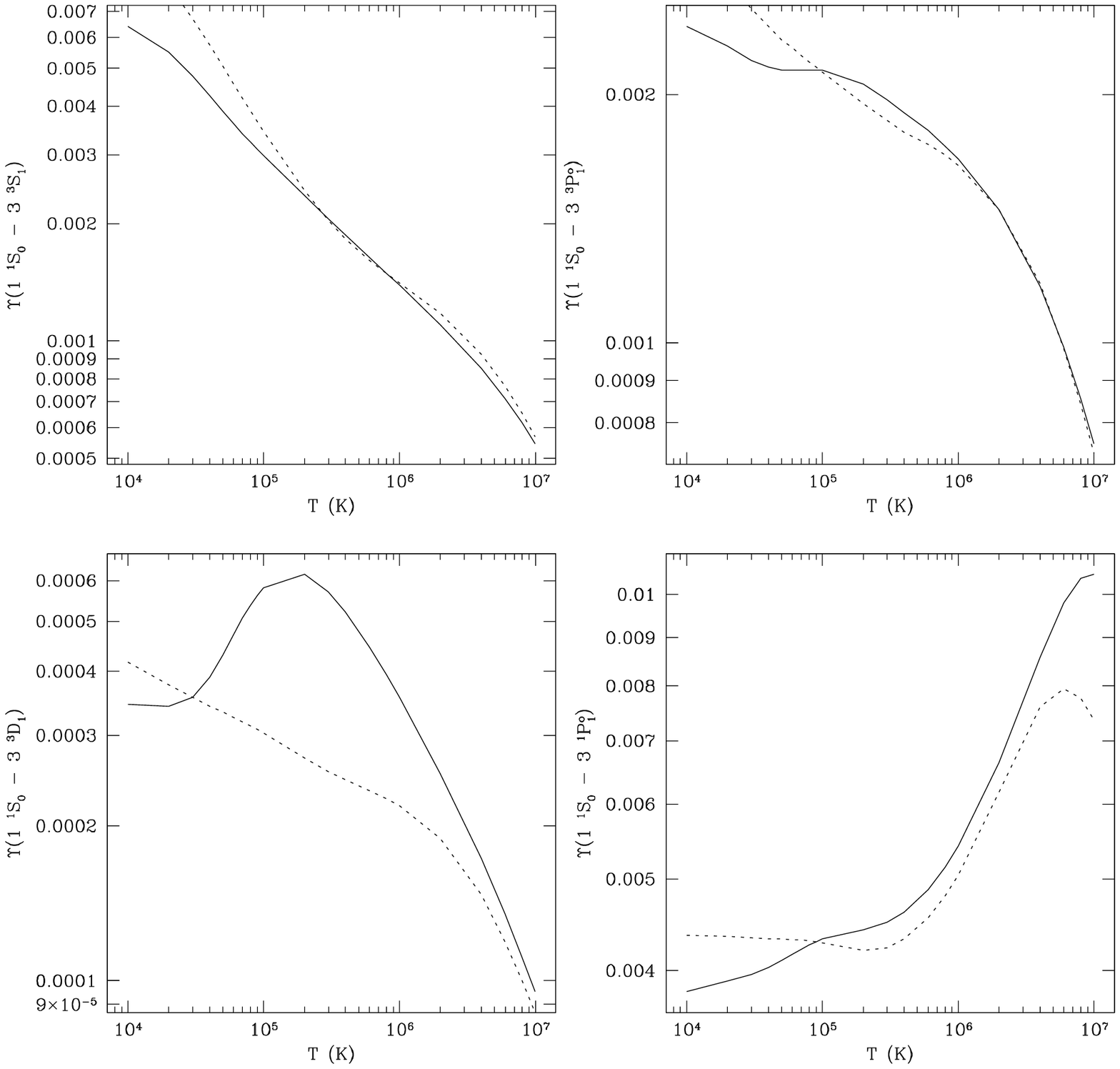,height=17.0cm,width=17.0cm} 
\caption{Effective Collision Strengths for transitions from $1s^2\ ^1S_0$ to 
the $n=3$ complex. 
solid line: Present work up to $n=4$ 
dashed line: Present work up to $n=3$ 
Note: The effect of resonance arising from $n=4$ complex may have a  
major effect 
on transition 
to the $n=3$ complex (a factor 2 for 1 $^1S_0 - 3 ^3D_1$)} 
\end{figure} 
 

 We compared the effective collision strengths with previous calculations
 for the principal lines w, x, y, z, (figure 6a 6b and 6d)corresponding to the 
 4 transitions to
 the ground level $1s^2 \ (^1S_0) \longleftarrow 1s2p (^1P^o_1), 1s2p
(^3P^o_2), 1s2p (^3P^o_1), 1s2s (^3S_1)$ respectively, that are of primary
interest in X-ray spectral diagnostics (e.g. Gabriel and Jordan 1969,
Pradhan 1982, Porquet et. al. 2001). Some other transitions are also
compared for which the data are available in literature. Illustrative
results are
presented for the four principal lines (figure 6) as well as four other transitions
within the $n=2$ complex (figure 7).

 Generally the agreement between the different calculations, depending on the
transition and temperature, is between 10 - 30\%. However, for some transitions
(including the important z-line transition) the differences are larger
in some temperature ranges. Basically, theses differences stem from
(i) the coupling effects due to the $n=3$ and $n=4$ levels in the close
coupling expansion, (ii) relativistic
effects included through the Breit-Pauli approximation, (iii)
improved delineation of resonances with high-resolution, and (iv)
ensuring convergence with complete `top-up' of partial waves. All of
these four factors are important in determining the final effective
collision strength.

 In figure 6 we present the transition from the ground level to the
different fine structure levels with $n=2$.
 The z-line (figure 6a) presents the biggest difference with previous 
calculations,
especially at low temperatures. There is a factor 2 difference
with Pradhan \etal (1981, crosses), and between 30 \% and a factor of 2 with 
Kinston 
and Tayal (1983, dot-dashes) for $T < 2\ 10^{5}$ K, both lying below the
present values. Above $10^{5}$ K the difference is within 10\%.
 The most striking difference however is with the results of
Zhang and Sampson (1987, short-dashes) that are much lower and start to
converge only at very high temperatures. One might ascribe it to the
fact that resonances are not fully considered in their work.

 The primary cause of differences with previous works is the resolution of
complexes of resonances near threshold, {\it in between the $n=2$ levels},
where high resolution
is crucial and  detailed fine structure plays an important role. The
$\Omega_{ij}$ at low energies determines the effective collision
strength at low temperatures since
$\Upsilon _{ij} (T \rightarrow 0) = \Omega _{ij} (E \rightarrow 0) $.
The importance of the resonances for this transition was already shown by
Pradhan \etal (1981) and Kinston and Tayal (1983a,b) in their LS coupling
calculations, both of whom obtain results significantly higher than
Zhang and Sampson (1987). The present results demonstrate the need for a 
relativistic and high-resolution calculation in order to accurately 
obtain the excitation rates at low temperatures.

 With the exception of Zhang and Sampson (1987), the previous works 
did not consider fine structure. Therefore
 the effective collision strengths for the $1^1S_0 - 2^3P^o_{2,1}$
transitions, corresponding to the forbidden and intercombination lines x
and y, are shown individually in figure 6b, as well as added together
$\Upsilon (x+y)$. The difference with all others
is no more than 20\% at all temperatures.

 For the dipole allowed transition ($1\ ^1S_0\ -\ 2\ ^1P^o$),
corresponding to the w-line,
the high temperature rates are within 10\%. But still, the low temperature
rates are higher by 20\% than those from Pradhan et. al. (1981, crosses).

 The strong dipole transitions among the excited $n=2$ levels
($2\ ^3S_1\ -\ 2\ ^3P^o$ and $2\ ^1S_0\ -\ 2\ ^1P^o$)  (figures 7c and 7d) are
very important in spectral diagnostics calculations since they enable
the collisional coupling at high electron densities that affects the
x, y, and z lines. This also has implications for the competition
between the effect of collisional redistribution among these lines, and
photoexcitation by background ultraviolet radiation, if present in the
X-ray source (e.g. Porquet \etal 2001). The total multiplet collision strength 
for this transition differs from earlier works by up to 30\%.

\section{Conclusion\protect\\} 
 Some of the general conclusions of the paper are as follows.

 1. The most complete close coupling calculation using the Breit-Pauli
R-Matrix method has been 
carried out for helium-like oxygen, including resonances up to $n = 4$
levels.
Detailed studies of radiation damping indicate that it may have a 
significant effect on the detailed collision strengths in a small energy
region below the threshold(s) of convergence, but not on the effective 
collision strengths.
However, radiation damping is important for higher-Z elements 
since the transition probabilities increase with Z (Pradhan 1983a,b has
found the effect on the z-transition to be 9\% in $\Upsilon$ for Fe~XXV
at the temperature of maximum abundance of helium-like iron).

2. It is verified that the effects of coupling and resonances from the $n = N+1$
complex play an important role  
in effective collision strengths for the transitions to the complex
$n = N$.

3. The new results for the important z-line transition should significantly
affect the analysis of
O~VII X-ray spectra from photoionized sources (e.g. active galactic
nuclei), where O~VII may be abundant at relatively low temperatures.
In collisional ionized (coronal) sources the new
results may not affect the theoretically computed line intensities
significantly at temperatures close to maximum abundance, but should still
do so at lower temperatures.
It would be preferable to employ the present data in future
collisional-radiative and photoionization models.

 4. As all relevant atomic effects in electron-ion collisions  have been
considered, and resonances have been carefully delineated, we should
expect the present results to be of definitive accuracy. Nonetheless, we
conservatively estimate the precision to be about 10-20\%.

 5. All data will be electronically available from the first author from
delahaye@astronomy.ohio-state.edu.

 The authors would like to thank Dr. Werner Eissner for immense help
with many aspects of this work. We also thank Guo-Xin Chen for the
new and efficient version of STGF including radiation damping
used in these calculations.
This work was supported partially by the U.S. National Science Foundation
and by NASA.
The computational work was carried out on the massively 
parallel Cray T3E and the vector processor
Cray T94 at the Ohio Supercomputer Center in Columbus, Ohio.

 
 
\section*{References} 
 
\def\amp{{\it Adv. At. Molec. Phys.}\ } 
\def\apj{{\it Astrophys. J.}\ } 
\def\apjs{{\it Astrophys. J. Suppl. Ser.}\ } 
\def\apjl{{\it Astrophys. J. (Letters)}\ } 
\def\aj{{\it Astron. J.}\ } 
\def\aa{{\it Astron. Astrophys.}\ }
\def\aas{{\it Astron. Astrophys. Suppl.}\ } 
\def\aasup{{\it Astron. Astrophys. Suppl.}\ } 
\def\adndt{{\it At. Data Nucl. Data Tables}\ } 
\def\cpc{{\it Comput. Phys. Commun.}\ } 
\def\jqsrt{{\it J. Quant. Spectrosc. Radiat. Transfer}\ } 
\def\jpb{{\it Journal Of Physics B}\ } 
\def\pasp{{\it Pub. Astron. Soc. Pacific}\ } 
\def\mn{{\it Mon. Not. R. astr. Soc.}\ } 
\def\pra{{\it Physical Review A}\ } 
\def\prl{{\it Physical Review Letters}\ } 
\def\zpds{{\it Z. Phys. D Suppl.}\ } 
 
\begin{harvard} 
\item{}Bell R.H., Seaton M.J., 1985, AdSpR 15, 37 
 
\item{}Berrington K.A., Burke P.G., Chang J.J., \etal, 1974, Comput. 
Phys. Commun. 8, 149 
 
\item{}Berrington K.A., Burke P.G., Le Dourneuf M., \etal, 1978, Comput. 
Phys. Commun. 14, 367 
 
\item{}Berrington K.A., Eissner, W. and Norrington, P.H. 1995 \cpc 92 290 

\item{}Burke P.G., Seaton M.J., 1971, Math. Comput. Phys. 10, 1 
 
\item{}Burke P.G., Hibbert A., Robb W.D., 1971, J. Phys. B 4, 153 
 
\item{}Burke, V. M. and Seaton, M. J. 1986 \jpb 19 L533  
 
\item{}Eissner W., Jones M., Nussbaumer H., 1974, Comput. Phys. Commun. 8,270 

\item{}Kaastra J.S., Mewe, R., Liedahl, D.A., Komosa, S., and Brinkman, A.C.
2000 \aa 354 L83
 
\item{}Kingston A.E., Tayal S.S., 1983, J. Phys. B 16, 3465 
 
\item{}Kingston A.E., Tayal S.S., 1983, J. Phys. B 16, L53 
 
\item{}Porquet D, Mewe R, Dubau J, Raassen A J J, and Kaastra J S
2001 \aa 376 1113

\item{}Porquet D and Dubau J 2000 \aas 143 495

\item{}Pradhan A.K., 1981 \prl 47 79

\item{}Pradhan A.K., 1982 \apj 263 477 

\item{}Pradhan A.K., 1983 \pra 28 2113 (a) 2128 (b)
 
\item{}Pradhan A.K., Seaton M.J., 1985, J. Phys. B 18, 1631 
 
\item{}Pradhan A.K., Norcross D.W., Hummer D.G., 1981, Ap. J. 246, 1031 
 
\item{}Pradhan A.K., Norcross D.W., Hummer D.G., 1981, Phys. Rev. A 23,619 
 
\item{}Presnyakov and Urnov A.M., 1979, J. Phys. B 8, 1280 
 
\item{}Sampson D.H., Goett S.J., Clark R.E.H., 1983, Atomic Data Nuclear 
Data Tables 29,467 
 
\item{}Zhang H.L., Sampson D.H., 1987, Ap. J. Supp. Ser. 63, 487 
\end{harvard} 
 
\enddocument